\documentclass[twocolumn, showpacs,aps,superscriptaddress,prl]{revtex4}

\usepackage{graphicx}
\usepackage{dcolumn}
\usepackage{bm}

\newcommand{\dd}{\mathrm{d}}

\newcommand{\eq}[1]{(\ref{#1})}
\newcommand{\bun}{\hat{\mathbf{b}}}
\newcommand{\nun}{\hat{\mathbf{n}}}

\newcommand{\bv}{\mathbf{v}}

\newcommand{\bx}{\mathbf{x}}

\newcommand{\bB}{\mathbf{B}}
\newcommand{\bE}{\mathbf{E}}

\newcommand{\dotcross}{ \raise 0.65ex\hbox{${\scriptstyle {{_{\displaystyle \cdot}}\atop\times}}$} }
\newcommand{\crossdot}{ \raise 0.5ex\hbox{${\scriptstyle {{_\times}\atop{\displaystyle \cdot}}}$} }

\newcommand{\sumsig}{ \raise -1.3ex\hbox{${{\displaystyle \sum}\atop{\scriptstyle \sigma}}$} }
\newcounter{appnumb}

\begin{document}


\title{When omnigeneity fails}

\author{F.~I. Parra}
\email{f.parradiaz1@physics.ox.ac.uk}
\affiliation{Rudolf Peierls
  Centre for Theoretical Physics, University of Oxford, Oxford, OX1
  3NP, UK} 
\affiliation{Culham Centre for Fusion Energy, Abingdon, OX14 3DB, UK}
\author{I. Calvo} 
\affiliation{Laboratorio
    Nacional de Fusi\'on, CIEMAT, 28040 Madrid, Spain}
\author{J.~L. Velasco}
\affiliation{Laboratorio Nacional de Fusi\'on, CIEMAT,
    28040 Madrid, Spain} 
\author{J.~A. Alonso}
\affiliation{Laboratorio Nacional de Fusi\'on, CIEMAT,
    28040 Madrid, Spain}

\date{\today}

\begin{abstract}
A generic non-symmetric magnetic field does not confine magnetized charged particles for long times due to secular magnetic drifts. Stellarator magnetic fields should be omnigeneous (that is, designed such that the secular drifts vanish), but perfect omnigeneity is technically impossible. There always are small deviations from omnigeneity that necessarily have large gradients.  The amplification of the energy flux caused by a deviation of size $\epsilon$ is calculated and it is shown that the scaling with $\epsilon$ of the amplification factor can be as large as linear. In opposition to common wisdom, most of the transport is not due to particles trapped in ripple wells, but to the perturbed motion of particles trapped in the omnigeneous magnetic wells around their bounce points. 
\end{abstract}

\maketitle

\emph{Introduction.}  Stellarators are non-axisymmetric magnetic
confinement devices used in fusion research. Unlike in
  axisymmetric tokamaks, the stellarator magnetic field is created
  only by external magnets, without the need of any
  mechanism to drive current within the plasma, thus reducing capital
  costs, providing a solution to the continuous operation required for
  a fusion reactor, and preventing some virulent macroscopic
  instabilities \cite{helander12}.

The magnetic field in a stellarator needs to be
  non-axisymmetric to form nested toroidal surfaces that hold the hot
  fusion plasma. In general, particles are not perfectly confined by
  magnetic fields without any continuous symmetry. They are only
confined to lowest order in $\rho_* = \rho_i/L \ll
  1$, where $\rho_i$ is the characteristic ion
gyroradius and $L$ is a characteristic length of the
problem. To next order in $\rho_*$, for magnetic
fields without any symmetry, particles drift away from magnetic field
lines secularly. These secular drifts lead
to large displacements and dominate the particle and energy losses in
stellarators.

  The transport due to large secular drifts can be reduced with a wise
  design \cite{wobig93, nuhrenberg95, anderson95, neilson02}. Ideally,
  stellarators would achieve perfect omnigeneity, that is, the average
  drift out of the core of the stellarator would be
  exactly zero. One of the main design objectives of the large stellarator Wendelstein 7-X is to be as close to omnigeneity as is technically possible \cite{wobig93, nuhrenberg95}. Cary and Shasharina \cite{cary97a,
    cary97b} showed that perfectly omnigeneous magnetic fields with
  continuous derivatives to all orders do not exist, but they rightly
  argued that this mathematical constraint does not preclude the
  possibility of reducing the transport due to large secular drifts
  considerably. If one assumes that the magnetic field
    and all its derivatives are continuous, omnigeneity is equivalent
    to a more restrictive condition on the magnetic field called
    quasisymmetry \cite{cary97a, cary97b}, and quasisymmetry is
    impossible to achieve in non-axisymmetric toroidal configurations
    \cite{garren91b}. However, it is possible to get very close to
    omnigeneity and yet be far from quasisymmetry. If we have a
    magnetic field that is omnigeneous but does not have continuous
    second or third derivatives, there always is a magnetic field with
    all its derivatives continuous that is as close as desired to the
    omnigeneous magnetic field \cite{cary97a, cary97b}. The
    non-omnigeneous part of the magnetic field will tend to have large
    higher order derivatives because it tries to be close to the
    discontinuous behavior of the perfecty omnigeneous magnetic
    field. Technically, getting arbitrarily close to
  omnigeneity can be prohibitively expensive because it requires large currents to ensure penetration of all the large helicity components of the magnetic field, and very precise alignment of these currents.

\begin{figure}

\includegraphics{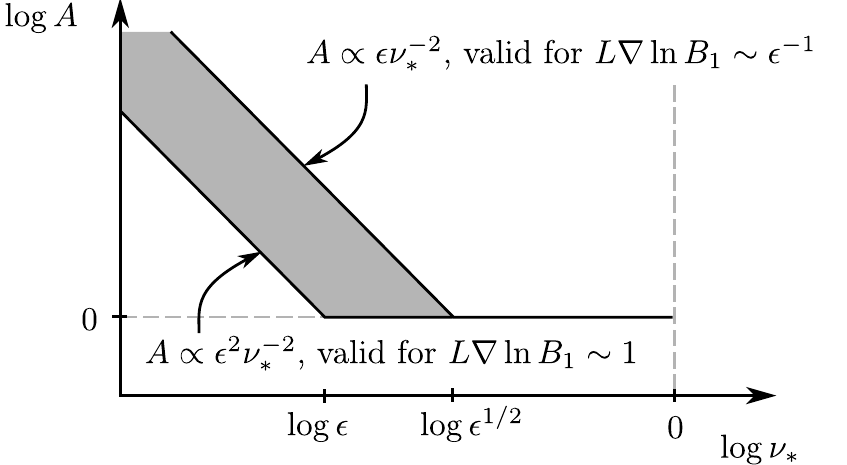}

\caption{ \label{fig:amplification} Energy flux
    amplification due to deviations from omnigeneity as function of
  $\nu_\ast$. When the deviation from omnigeneity gives flux smaller
  than the flux of a perfectly omnigeneous stellarator, we set the
  amplification to $1$.}

\end{figure}

  In this letter, we study what unavoidable small deviations from
  omnigeneity do to ion energy transport (the same results apply to
  ion particle transport, or electron particle and energy
  transport). We calculate the amplification of the
    energy flux due to deviations from omnigeneity
  and identify its causes. In particular, we prove
    that the degradation of confinement is not dominated by ripple
    wells, as has often been assumed. Our results are summarized in
  Fig. \ref{fig:amplification}.

  \emph{Magnetic coordinates in stellarators.} The magnetic field in a
  stellarator forms nested toroidal surfaces known as flux
  surfaces. To locate a spatial point $\bx$, we use a radial variable
  $r (\bx)$ with dimensions of length that determines in which flux
  surface the point is, and two variables that determine the location
  of the point within the flux surface: the length along the magnetic
  field line, $l(\bx)$, and an angle $\alpha(\bx)$ that gives the
  position perpendicular to the magnetic field line within the flux
  surface. Inverting $r(\bx)$, $\alpha(\bx)$ and $l(\bx)$, we find
  $\bx (r, \alpha, l)$. The angle $\alpha$ is defined such that $\bB =
  \Psi_\zeta^\prime \nabla r \times \nabla \alpha$, where $\Psi_\zeta
  (r)$ is the toroidal magnetic flux enclosed by the flux surface $r$
  divided by $2\pi$, and $\Psi_\zeta^\prime = \dd \Psi_\zeta/\dd r$.

\emph{Equations for transport in stellarators.} In this letter we
calculate the radial energy flux $Q_i = \int \dd^2 S \int \dd^3v\, f_i
(m_i v^2/2) \bv\cdot \nun$, where $f_i (\bx, \bv)$ is the ion
distribution function, $m_i$ is the ion mass, $\nun =
  \nabla r/|\nabla r|$ is the normal to the flux surface, and
\begin{equation} \label{eq:d2S}
\int \dd^2 S\, (\ldots) = \Psi_\zeta^\prime \int_0^{2\pi} \dd \alpha \int_0^L \dd l \, \frac{|\nabla r|}{B} (\ldots)
\end{equation}
is the integral over the flux surface. The limit $L(r, \alpha)$ in the integral over $l$ depends on both $r$ and $\alpha$.

To calculate $Q_i$, we assume an ordering typical of hot stellarator
core plasmas, $\rho_\ast \ll \nu_\ast \ll 1$, where $\nu_\ast = L
\nu_{ii}/v_{ti}$, $\nu_{ii}$ is the ion-ion collision frequency,
$v_{ti} = \sqrt{2T_i/m_i}$ is the ion thermal speed, and $T_i$ is
the ion temperature. With
  this ordering, the ion distribution function is a stationary
  Maxwellian to lowest order in $\rho_\ast$, $f_{Mi} (r, v) = n_i (r)
  ( m_i/2\pi T_i(r) )^{3/2} \exp ( - m_i v^2/2 T_i(r) )$, where the
  ion density and temperature are constants within the flux
  surface. The electric field is electrostatic, $\bE = - \nabla \phi$,
  and to lowest order, due to quasineutrality, the electrostatic
  potential is constant within the flux surface, $\phi \simeq \phi_0
  (r)$. The lowest order potential satisfies $e\phi_0/T_i \sim 1$,
  where $e$ is the proton charge.

  The corrections to $f_{Mi}$ and $\phi_0$ are calculated by expanding
  first in $\rho_\ast \ll 1$ and later in $\nu_\ast \ll 1$. To lowest
  order, the three natural variables to describe the velocity are the
  magnitude of the velocity $v$, $\lambda = v_\bot^2/v^2 B$ and the
  gyrophase $\varphi$, which is the angle that gives the direction of
  the component of the velocity that is perpendicular
  to the magnetic field, $\bv_\bot$. In addition to $v$, $\lambda$ and
  $\varphi$, it is necessary to specify the sign of the parallel
  velocity, $\sigma = \pm 1$.

    We first expand in $\rho_\ast \ll 1$, finding $\phi = \phi_0 +
    \phi_1$ and $f_i = f_{Mi} + f_{i1}$, where $\phi_1 \sim \rho_\ast
    T_i/e$ and using the drift kinetic formalism \cite{hazeltine73},
    $f_{i1} = h_i - (Ze\phi_1/T_i) f_{Mi} + (\Upsilon_i
    f_{Mi}/\Omega_i) (\bv \times \bun) \cdot \nabla r$. The function
    $h_i (r, \alpha, l, v, \lambda, \sigma) \sim \rho_\ast f_{Mi}$ is
    independent of the gyrophase. Here $\Upsilon_i = \partial_r \ln
    n_i + Ze \partial_r \phi_0/T_i+ ( m_i v^2/2 T_i - 3/2 ) \partial_r
    \ln T_i$, $\Omega_i = ZeB/m_i c$ is the ion gyrofrequency, $\bun =
    \bB/B$ is the unit vector in the direction of the magnetic field,
    $Ze$ the ion charge, and $c$ the speed of light. The equation for
    $h_i (r, \alpha, l, v, \lambda, \sigma)$ is
    \begin{equation} \label{eq:FPorder1} v_{||} \partial_l h_i +
      \Upsilon_i f_{Mi} v_{Mi, r} |\nabla r| = C_{ii}^{(\ell)} [ h_i
      ],
\end{equation}
where $v_{||} = \sigma v \sqrt{1 - \lambda B(r, \alpha, l)}$,
\begin{equation} \label{eq:vMir} v_{Mi,r} = \frac{v^2(2 - \lambda B)}{2B\Omega_i |\nabla r|}
    (\bun \times \nabla B) \cdot \nabla r
\end{equation}
is the radial magnetic drift, and $C_{ii}^{(\ell)} [h_i]$ is the
linearized Fokker-Planck collision operator. The operator
$C_{ii}^{(\ell)}$ represents the collisions with the background
Maxwellian. It is a linear integro-differential operator with
coefficients that only depend on $\alpha$ and $l$ via the magnitude of
the magnetic field $B(r, \alpha, l)$ that enters in the collision
operator because of the definition of $\lambda$.

For $\rho_\ast \ll \nu_\ast \ll 1$, the energy flux becomes 
\begin{equation} \label{eq:Qifinal} Q_i = \int \dd^2 S \int \dd^3v\,
  h_i \frac{m_i v^2}{2} v_{Mi, r} + O(\nu_\ast \rho_\ast^2 p_i v_{ti}
  S_r),
\end{equation}
where $p_i = n_i T_i$ is the ion pressure and $S_r = \Psi_\zeta^\prime
\int_0^{2\pi} \dd \alpha \int_0^L \dd l\, |\nabla r|/B$ is the area of
the flux surface. The term of order $\nu_\ast \rho_\ast^2 p_i v_{ti}
S_r$ is important in the perfectly omnigeneous case, but
in this letter we do not need to know its exact
form. The velocity integral written in the variables $v$, $\lambda$
and $\varphi$ gives
\begin{equation} \label{eq:intv} \int \dd^3v\, (\ldots) = \sum_\sigma
  \int_0^\infty \dd v \int_0^{B^{-1}} \dd \lambda\, \frac{\pi B
    v^3}{|v_{||}|} ( \ldots ).
\end{equation}
In this equation, the summation sign $\sum_\sigma$ indicates that we
have to sum over both signs of the parallel velocity, $\sigma = +1$
and $\sigma = -1$.

\emph{Perfectly omnigeneous stellarators.} To lowest order in a
subsidiary expansion in $\nu_\ast \ll 1$, equation \eq{eq:FPorder1}
becomes $\partial_l h_i = 0$. Trapped particles ($\lambda >
B^{-1}_\mathrm{max}$, where $B_\mathrm{max} (r)$ is the maximum value
of $B(r, l, \alpha)$) have bounce points at $l = l_b$, where $v_{||} =
\sigma v \sqrt{1 - \lambda B}$ vanishes because $B (r, \alpha, l_b) =
\lambda^{-1}$. At $l = l_b$, $h_i (\sigma = -1) = h_i (\sigma = -1)$,
and $\partial_l h_i=0$ therefore implies that for
trapped particles, $h_i (r, \alpha, v, \lambda)$ does
  not depend on $l$ or $\sigma$. For passing particles ($\lambda <
  B_\mathrm{max}^{-1}$), $v_{||}$ never goes to zero, and in an
  ergodic flux surface, a passing particle samples the entire flux
  surface by moving along the magnetic field line. As a result,
  $\partial_l h_i = 0$ implies that for passing particles, $h_i (r, v,
  \lambda, \sigma)$ does not depend on $\alpha$ in addition to not
  depending on $l$. Passing particles in rational flux surfaces where
  the magnetic field lines close on themselves can be treated as
  trapped particles. To summarize, in an ergodic flux surface, $h_i
  (r, \alpha, v, \mu, \sigma) = H_i (r, v, \mu, \sigma) + h_{i,t} (r,
  \alpha, v, \mu)$, where $h_{i,t}$ is non-zero only in the trapped
  region $\lambda > B^{-1}_\mathrm{max}$. By continuity, $h_{i,t} = 0$
  at the boundary between trapped and passing particles, $\lambda =
  B^{-1}_\mathrm{max}$. To completely define $h_{i,t}$, we impose that
  $\int_0^{2\pi} \dd\alpha\, h_{i,t} = 0$.

\begin{figure}

\includegraphics{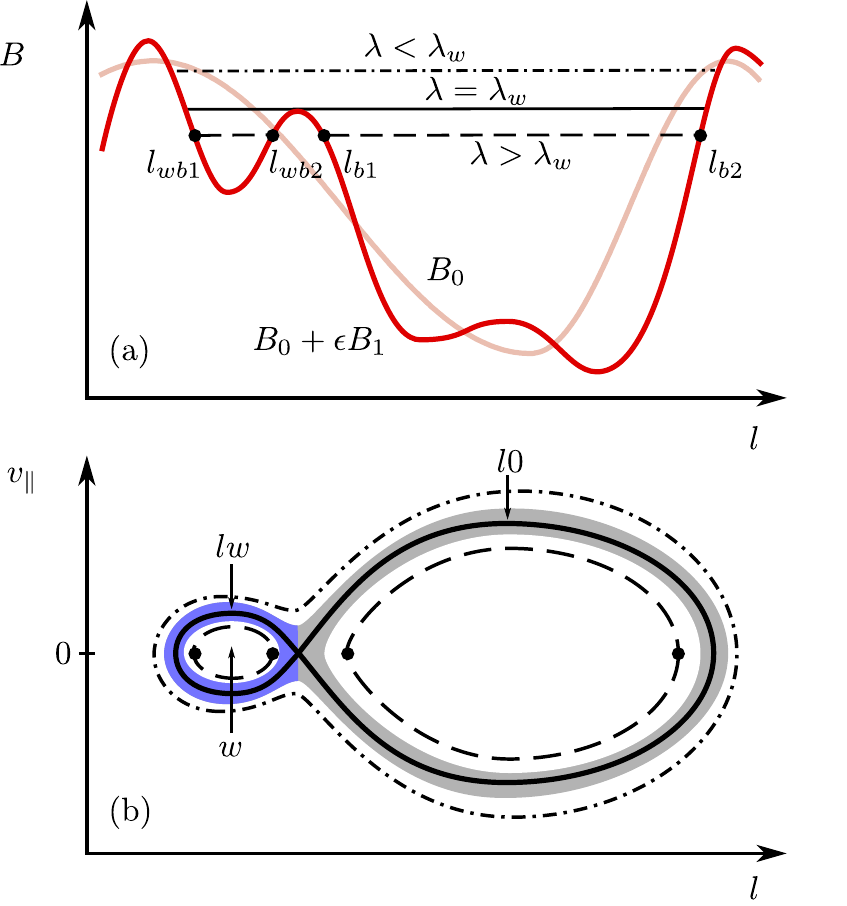}

\caption{ \label{fig:Bsketch} (a) Omnigeneous magnetic field $B_0$ as a function of $l$ (light), plus deviation from omnigeneity $\epsilon B_1$ (dark). (b) Particle orbits for $B = B_0 + \epsilon B_1$ in $v_{||}$ vs. $l$ space.}

\end{figure}

  To obtain equations for $h_{i,t}$ and $H_i$, we eliminate the term
  $v_{||} \partial_l h_i$ in \eq{eq:FPorder1} by integrating over
  orbits for trapped
    particles, $\lambda > B_\mathrm{max}^{-1}$, and by integrating
    equation \eq{eq:FPorder1} multiplied by $B/|v_{||}| |\nabla r|$
    over the entire flux surface for passing particles, $\lambda <
    B^{-1}_\mathrm{max}$, leaving
\begin{equation} \label{eq:FPtrapped}
\sum_{\sigma} \int_{l_{b1}}^{l_{b2}} \frac{\dd l}{|v_{||}|} C_{ii}^{(\ell)} [h_{i,t} + H_i] = \frac{m_i c}{Ze\Psi^\prime_\zeta} \Upsilon_i f_{Mi} \partial_\alpha J
\end{equation}
for $\lambda > B_\mathrm{max}^{-1}$, and
\begin{equation} \label{eq:FPpassing}
\int \dd^2 S \frac{B}{|v_{||}| |\nabla r|} C_{ii}^{(\ell)} [h_{i,t} + H_i] = 0
\end{equation}
for $\lambda < B_\mathrm{max}^{-1}$. Here $J = 2\int_{l_{b1}}^{l_{b2}} \dd l\, |v_{||}|$ is the second adiabatic invariant \cite{northrop63}, and $l_{b1}$ and $l_{b2}$ are the bounce points, that is, $B(r, \alpha, l_{b1}) = B(r, \alpha, l_{b2}) = \lambda^{-1}$ (see Fig.~\ref{fig:Bsketch}(a)). To obtain the right side of these equations, we have used the well known results \cite{boozer04}
\begin{equation} \label{eq:averagevMir}
\sum_{\sigma} \int_{l_{b1}}^{l_{b2}} \frac{\dd l}{|v_{||}|} v_{Mi, r}|\nabla r| = - \frac{m_i c}{Ze \Psi^\prime_\zeta} \partial_\alpha J
\end{equation} 
and $\int \dd^2 S (B/|v_{||}|) v_{Mi,r} = 0$.

If $\partial_\alpha J = 0$, both $h_{i,t}$ and $H_i$ vanish, and we
need to go to next order in $\nu_\ast$, making fluxes such as
\eq{eq:Qifinal} small in $\nu_\ast \ll 1$. That is,
  perfect omnigeneity is achieved when $J = 2v \int_{l_{b1}}^{l_{b2}}
  \dd l\, \sqrt{1 - \lambda B (r, \alpha, l)}$ does not depend on
  $\alpha$. Since this has to be satisfied for every $\lambda$, the
  final condition is that \cite{cary97a, cary97b}
\begin{equation} \label{eq:omni}
\partial_\alpha \left [ \int_{l_{b1}}^{l_{b2}} \dd l\, F(r, v, \lambda, B(r, \alpha, l)) \right ] = 0
\end{equation}
for any function $F$ that only depends on $\alpha$ and $l$ via the
magnitude of the magnetic field magnitude $B(r, \alpha, l)$. This
condition constrains how $B$ depends on $l$.

As explained in the introduction, it is technically
  impossible to achieve perfectly omnigenous fields, but it is feasible
  to get close to omnigeneity. To treat deviations from omnigeneity,
we consider $\bB = (B_0 + \epsilon B_1) \bun$, with $\epsilon \ll 1$,
$B_0$ the omnigeneous magnetic field, and $B_1$ the non-omnigeneous
part. Since we expect $B_1$ to have large derivatives, we consider
both $L \nabla \ln B_1 \sim \epsilon^{-1}$ and $L \nabla \ln B_1 \sim
1$ to bound the effect of deviations from omnigeneity. It is
convenient to start by assuming $L \nabla \ln B_1 \sim \epsilon^{-1}$,
and then take the limit $L \nabla \ln B_1 \sim 1$ as a subsidiary
expansion. We will compare the energy flux due to deviations from
omnigeneity with the energy flux in a perfectly omnigeneous
stellarator, given in order of magnitude by \cite{landreman12}
\begin{equation} \label{eq:Qiomni}
Q_i^\mathrm{om} = O ( \nu_\ast \rho_\ast^2 p_i v_{ti} S_r).
\end{equation}

\emph{Perturbation to omnigeneity with large gradients.} We assume $L \nabla \ln B_1 \sim \epsilon^{-1}$. Equation \eq{eq:vMir} implies that $v_{Mi,r}$ is close to the perfectly omnigeneous radial magnetic drift, $v_{Mi,r}^{(0)}$, only if $\epsilon (\bun \times \nabla B_1) \cdot \nabla r \ll (\bun \times \nabla B_0) \cdot \nabla r$. It is sufficient if $\epsilon (\bun \times \nabla B_1) \cdot \nabla r = O ( \epsilon^{1/2} B_0/L ) \ll (\bun \times \nabla B_0) \cdot \nabla r$, that is, in a stellarator close to omnigeneity, the angle between $\bun$ and the component of $\nabla B_1$ parallel to the flux surface is of the order of or smaller than $\epsilon^{1/2}$. This assumption can be written as
\begin{equation} \label{eq:angleB1b}
(\bun \cdot \partial_\alpha \bx) \partial_l B_1 - \partial_\alpha B_1 = O ( \epsilon^{-1/2} B_0 ).
\end{equation}

Assuming that \eq{eq:angleB1b} is satisfied, we can expand \eq{eq:FPorder1} in $\epsilon^{1/2}$. To lowest order, we can replace $B$ by $B_0$ in the terms on the left side of equations \eq{eq:FPtrapped} and \eq{eq:FPpassing}. We cannot do that for the right side of \eq{eq:FPtrapped}, as we will see. We use the superindex $^{(0)}$ to indicate that $B$ has been replaced by $B_0$. According to \eq{eq:omni}, the coefficients of the operator $\int_{l_{b1}^{(0)}}^{l_{b2}^{(0)}} \dd l\, (|v_{||}|^{(0)})^{-1} (C_{ii}^{(\ell)})^{(0)}$ are independent of $\alpha$, and as a result, the effect of collisions between trapped and passing particles averages to zero when all trapped particles are considered (recall that $\int_0^{2\pi} \dd \alpha\, h_{i,t} = 0$). Then, $H_i$ is zero to lowest order, and we are only left with $h_{i,t}$, determined by
\begin{equation} \label{eq:FPtrappedfinal}
\sum_{\sigma} \int_{l_{b1}^{(0)}}^{l_{b2}^{(0)}} \frac{\dd l}{|v_{||}|^{(0)}} (C_{ii}^{(\ell)})^{(0)} [h_{i,t}] = \frac{m_i c}{Ze\Psi^\prime_\zeta} \Upsilon_i f_{Mi} (\partial_\alpha J)^{(1)},
\end{equation}
where $(\partial_\alpha J)^{(1)} \sim \epsilon^{1/2} v_{ti} L$. This equation leads to $h_{i,t} \sim \epsilon^{1/2} \nu_\ast^{-1} \rho_\ast f_{Mi}$. To prove that $\partial_\alpha J \simeq (\partial_\alpha J)^{(1)} \sim \epsilon^{1/2} v_{ti} L$, $J$ must be expanded in $\epsilon^{1/2}$ as $J = J^{(0)} + J^{(2)} + J^{(3)} + \ldots$, where $J^{(0)} = 2v \int_{l_{b1}}^{l_{b2}} \dd l\, \sqrt{1 - \lambda B_0}$. The next order corrections $J^{(2)} \sim \epsilon v_{ti} L$ and $J^{(3)} \sim \epsilon^{3/2} v_{ti} L$, are obtained by splitting the integral between $l_{b1}$ and $l_{b2}$ into three different regions: $[l_{b1}, l_{b1} + \Delta l_{b1}]$, $[l_{b1} + \Delta l_{b1}, l_{b2} - \Delta l_{b2}]$ and $[l_{b2} - \Delta l_{b2}, l_{b2}]$, where $\Delta l_{b1} \sim \epsilon L$ and $\Delta l_{b2} \sim \epsilon L$ are chosen such that $1 - \lambda B_0 \gg \lambda \epsilon B_1$ for $l \in [l_{b1} + \Delta l_{b1}, l_{b2} - \Delta l_{b2}]$. The correction $J^{(2)}$ is the correction to the integral over the region $[l_{b1} + \Delta l_{b1}, l_{b2} - \Delta l_{b2}]$, where we can Taylor expand $B_0 + \epsilon B_1$ around $B_0$ to find the first order correction
\begin{equation} \label{eq:J2}
J^{(2)} = - \epsilon v \lambda \int_{l_{b1} + \Delta l_{b1}}^{l_{b2} - \Delta l_{b2}} \dd l\, \frac{B_1}{\sqrt{1 - \lambda B_0}} = O(\epsilon v_{ti} L).
\end{equation} 
In the integrals over $[l_{b1}, l_{b1} + \Delta l_{b1}]$ and $[l_{b2} - \Delta l_{b2}, l_{b2}]$, we cannot Taylor expand because $1 - \lambda B_0 \sim \lambda \epsilon B_1$,
\begin{eqnarray} \label{eq:J3}
J^{(3)} = 2v \int_{l_{b1}}^{l_{b1} + \Delta l_{b1}} \dd l\, (\sqrt{1 - \lambda B_0 - \lambda \epsilon B_1} - \sqrt{1 - \lambda B_0}) \nonumber\\ + 2v \int_{l_{b2} - \Delta l_{b2}}^{l_{b2}} \dd l\, (\sqrt{1 - \lambda B_0 - \lambda \epsilon B_1} - \sqrt{1 - \lambda B_0}).
\end{eqnarray}
Since $\sqrt{1 - \lambda B_0 - \epsilon B_1} \sim \sqrt{1 - \lambda B_0} \sim \epsilon^{1/2}$ over a length $O(\epsilon L)$, $J^{(3)} \sim \epsilon^{3/2} v_{ti} L$. For \eq{eq:FPtrappedfinal} we need $\partial_\alpha J$. Because $B_0$ is omnigeneous, $\partial_\alpha J^{(0)} = 0$. Then, $\partial_\alpha J \simeq (\partial_\alpha J)^{(1)} = \partial_\alpha J^{(2)} + \partial_\alpha J^{(3)} \sim \epsilon^{1/2} v_{ti} L$. The term $\partial_\alpha J^{(3)}$ is of order $\epsilon^{1/2} v_{ti} L$ because $J^{(3)} \sim \epsilon^{3/2} v_{ti} L$ and $\partial_\alpha \ln B_1 \sim \epsilon^{-1}$. To prove that $\partial_\alpha J^{(2)} \sim \epsilon^{1/2} v_{ti} L$, we take the derivative with respect to $\alpha$ of \eq{eq:J2}, we use \eq{eq:angleB1b} to write $\partial_\alpha B_1 = (\bun \cdot \partial_\alpha \bx) \partial_l B_1 + O(\epsilon^{1/2} B_0)$, and we integrate by parts in $l$ to find $\partial_\alpha J^{(2)} \sim \epsilon^{1/2} v_{ti} L$. Our assumption \eq{eq:angleB1b} was crucial to show that $\partial_\alpha J^{(2)} \sim \epsilon^{1/2} v_{ti} L$. Relation \eq{eq:angleB1b} must be the objective of stellarator design because our estimate of $J^{(3)}$ in \eq{eq:J3} necessarily gives $\partial_\alpha J^{(3)} \sim \epsilon^{1/2} v_{ti} L$, and reducing $(\bun \times \nabla B_1) \cdot \nabla r$ further than assumed in \eq{eq:angleB1b} is not worthwhile. 

With \eq{eq:d2S} and \eq{eq:intv}, we can calculate the energy flux \eq{eq:Qifinal},
\begin{equation} \label{eq:Qieps}
Q_i = \frac{\pi m_i^2 c}{2Ze} \int_0^\infty \dd v \int_{B_\mathrm{max}^{-1}}^{B_\mathrm{min}^{-1}} \dd \lambda \int_0^{2\pi} \dd \alpha\, h_{i,t} (\partial_\alpha J)^{(1)} v^5,
\end{equation}
where we have used that $h_{i,t}$ does not depend on $l$ and $\sigma$, and that $h_{i,t}$ is non zero only for trapped particles, $\lambda \in [ B_\mathrm{max}^{-1}, B_\mathrm{min}^{-1}]$, where $B_\mathrm{min} (r)$ is the minimum value of $B(r, \alpha, l)$ in the flux surface $r$. We have also used \eq{eq:averagevMir} to simplify $\sum_{\sigma} \int_{l_{b1}}^{l_{b2}} \dd l\, v_{Mi, r}|\nabla r|/|v_{||}|$. From \eq{eq:Qieps}, it is obvious that
\begin{equation} \label{eq:Qiestimate}
Q_i = O( \epsilon \nu_\ast^{-1} \rho_\ast^2 p_i v_{ti} S_r ).
\end{equation}
This flux is larger than the omnigeneous flux \eq{eq:Qiomni} for $\epsilon^{-1/2} \nu_\ast \ll 1$, giving an amplification $A = Q_i/Q_i^\mathrm{omni} \sim \epsilon \nu_\ast^{-2}$. For $\epsilon^{-1/2} \nu_\ast \gg 1$, the omnigeneous flux \eq{eq:Qiomni} is dominant, and we need not worry about deviations from omnigeneity. 

\emph{Small ripple wells}. Small ripple wells like the ones shown in Fig.~\ref{fig:Bsketch}(a) can form when $L \nabla \ln B_1 \sim \epsilon^{-1}$ because it is possible to have points at which $\partial_l B_0 + \epsilon \partial_l B_1 = 0$. The calculation so far has ignored these ripple wells. They turn out to be unimportant for the scaling.

Ripple wells affect three small regions in phase space, depicted in Fig.~\ref{fig:Bsketch}(b): the well $w$, and the layers $lw$ and $l0$. The characteristic size of these regions is given in Table~\ref{table:layers}. Ripple trapped particles in $w$ move across flux surfaces to get to the flux surface of interest. Via pitch angle diffusion, these particles can go into the layer $lw$, and moving along the magnetic field line, particles can then go into the layer $l0$. From both $lw$ and $l0$, particles can pitch angle scatter into other regions in phase space. As a result, there is a flux of particles leaving from $\lambda_w$, causing a discontinuity in the partial derivative $\partial_\lambda h_{i,t}$ at $\lambda_w$. By integrating in $l$ and $\lambda$ over regions $w$, $lw$ and $l0$, we can explicitly calculate the discontinuity in terms of the parameters of the ripple well \cite{parra14}. The size of the jump can be estimated from the characteristic $\Delta \lambda$ and $\Delta h_i$ of $l0$ (see Table~\ref{table:layers}). The jump in $\partial_\lambda h_{i,t}$ is of order $\epsilon \partial_\lambda h_{i,t}$. Even though this jump is small, in general we have a number of wells of order $\epsilon^{-1}$ in a given magnetic field line, and by accumulation, the effect of this jump condition modifies the distribution function $h_{i,t}$ by an amount of order one.

We now explain how to find the results in Table~\ref{table:layers} \cite{parra14}. The widths of the intervals in $\lambda$ are small (the width in $v_{||}$ of the regions sketched in Fig.~\ref{fig:Bsketch}(b) is $\Delta v_{||} \sim v_{||} B_0 \Delta \lambda$). As a result, $\partial_\lambda$ is large, and the pitch angle scattering piece dominates in the collision operator, $C_{ii}^{(\ell)} [ h_i ] \simeq (\nu_\lambda v_{||}/B v^2) \partial_\lambda \left ( \lambda v_{||} \partial_\lambda h_i \right )$. The frequency $\nu_\lambda (v)$ is the pitch angle scattering frequency. The width of the well $w$ is $\Delta \lambda \sim \epsilon/B_0$ because $v_{||} = \sigma v \sqrt{1 - \lambda B}$ must vanish for the range of values of $B$ in the ripple well. To estimate the change in the distribution function $\Delta h_i$ in the well $w$, we integrate an equation like \eq{eq:FPorder1} in the well. To find the widths $\Delta \lambda$ of the layers $lw$ and $l0$ we make collisions and parallel streaming comparable, and to estimate the changes in the distribution functions $\Delta h_i$ we impose continuity of derivatives between the well $w$ and the layer $lw$, and continuity of particle flow in phase space between the layers $lw$ and $l0$.

\begin{table}

\caption{\label{table:layers} Characteristic quantities in the regions in phase space that are affected by a small ripple well: parallel velocity $v_{||}$, length $\Delta l$, width of the interval in $\lambda$, $\Delta \lambda$, relative importance of collisions with respect to the parallel streaming $C_{ii}^{(\ell)} / v_{||} \partial_l$, and change of the distribution function within the layer $\Delta h_i$.}

\begin{tabular}{|c|c|c|c|c|c|}
\hline Region & $v_{||}/v_{ti}$ & $\Delta l/L$ & $B_0 \Delta \lambda$ & $C_{ii}^{(\ell)} / v_{||} \partial_l$ & $\Delta h_i/h_{i,t}$ \\
\hline Well $w$ & $\epsilon^{1/2}$ & $\epsilon$ & $\epsilon$ & $\epsilon^{-1/2} \nu_\ast$ & $\epsilon^{1/2}$ \\
\hline Layer $lw$ & $\epsilon^{1/2}$ & $\epsilon$ & $\epsilon^{3/4} \nu_\ast^{1/2}$ & $1$ & $\epsilon^{1/4} \nu_\ast^{1/2}$ \\
\hline Layer $l0$ & $1$ & $1$ & $\nu_\ast^{1/2}$ & $1$ & $\epsilon \nu_\ast^{1/2}$ \\
\hline
\end{tabular}

\end{table}

Note that the change in the distribution function across regions $w$, $lw$ and $l0$ are small compared to $h_{i,t}$. Then, in these regions $h_i$ is almost constant and of order $h_{i,t} \sim \epsilon^{1/2} \nu_\ast^{-1} \rho_\ast f_{Mi}$. The number of particles in the well and its surroundings is not controlled by the well itself, but by collisional balance with the particles trapped in the larger wells.

With these estimates, we can find the contribution from the ripple
wells to the energy flux. Using \eq{eq:d2S} and \eq{eq:intv}, the
contribution due to the region $w$ is $Q_{i,w} \sim B_0 \Delta \lambda
\Delta \alpha (\Delta l/L) (h_{i,t}/f_{Mi}) (v_{ti}/v_{||}) p_i
v_{Mi,r} S_r \sim \epsilon^3 \nu_\ast^{-1} \rho_\ast^2 p_i v_{ti}
S_r$, where $\Delta \alpha \sim \epsilon$ is the extent of the well in
$\alpha$, and we have neglected $\Delta h_i$ with respect to $h_{i,t}$. When $\epsilon^{-1/2} \nu_\ast \gg 1$, the small change $\Delta h_i$ in the layers $lw$ and $l0$
matters. The contributions from $lw$ and $l0$ are $Q_{i,lw} = O( \epsilon^3 \rho_\ast^2 p_i v_{ti} S_r
)$ and $Q_{i,l0} = O( \epsilon^{5/2} \rho_\ast^2 p_i v_{ti}
S_r)$. According to these estimates, $Q_{i,lw}$ is always negligible
because it is smaller than both $Q_{i,w}$ or $Q_{i,l0}$. For
$\epsilon^{-1/2} \nu_\ast \ll 1$, $Q_{i,w}$ is larger than $Q_{i,l0}$,
whereas for $\epsilon^{-1/2} \nu_\ast \gg 1$, $Q_{i,l0}$ is the
dominant contribution. The final effect of the ripple wells depends on
the total number of them. In general, we expect a number of ripple
wells of order $\epsilon^{-1}$ in each magnetic field line, and the
number of lines with ripple wells is of order $\epsilon^{-1}$, giving
a number of wells of order $\epsilon^{-2}$. Thus, for $\epsilon^{-1/2}
\nu_\ast \ll 1$, the total energy flux due to ripple wells is
$O(\epsilon \nu_\ast^{-1} \rho_\ast^2 p_i v_{ti} S_r)$, and for
$\epsilon^{-1/2} \nu_\ast \gg 1$, the flux due to ripple wells is
$O(\epsilon^{1/2} \rho_\ast^2 p_i v_{ti} S_r)$. If we compare these
estimates to the perfectly omnigeneous flux in \eq{eq:Qiomni}, we find
that the energy flux due to deviations from omnigeneity is higher than
the flux of a perfectly omnigeneous stellarator by $A \sim
\epsilon \nu_\ast^{-2}$ for $\epsilon^{-1/2} \nu_\ast \ll 1$, but it is
smaller for $\epsilon^{-1/2} \nu_\ast \gg 1$. These estimates are
exactly the same as the ones we obtained without ripple wells.

\emph{Perturbation to omnigeneity with small gradients.} To bound the
effect of deviations from omnigeneity, we consider $L \nabla \ln B_1
\sim 1$. We have evaluated the corrections to the second adiabatic
invariant due to deviations from omnigeneity, $J^{(2)}$ and $J^{(3)}$,
in \eq{eq:J2} and \eq{eq:J3}. For small gradients, the size of
$J^{(3)}$ is not $\epsilon^{3/2} v_{ti} L$. In the first integral in
\eq{eq:J3}, we can Taylor expand $B_0$ and $B_1$ around $l_{b1}$,
finding $\sqrt{1 - \lambda B - \epsilon \lambda B_1} \simeq \sqrt{-
  \lambda (\partial_l B_0 + \epsilon \partial_l B_1) (l - l_{b1})}$
and $\sqrt{1 - \lambda B_0} \simeq \sqrt{- \lambda \partial_l B_0 (l -
  l_{b1})}$. Similarly, in the second integral of \eq{eq:J3}, we can
Taylor expand $B_0$ and $B_1$ around $l_{b2}$. With these Taylor
expansions, it is easy to see that $J^{(3)}$ is of order $\epsilon^2
v_{ti} L$. Then, $\partial_\alpha J \simeq \partial_\alpha J^{(2)}$,
and since $\partial_\alpha \ln B_1 \sim 1$, $\partial_\alpha J \sim
\epsilon v_{ti} L$. As a result, an equation similar to
\eq{eq:FPtrappedfinal} gives $h_{i,t} \sim \epsilon \nu_\ast^{-1}
\rho_\ast f_{Mi}$, and equation \eq{eq:Qieps} leads to
\begin{equation} \label{eq:Qilowgradient} Q_i = O(\epsilon^2
  \nu_\ast^{-1} \rho_\ast^2 p_i v_{ti} S_r).
\end{equation}
Comparing \eq{eq:Qilowgradient} with \eq{eq:Qiomni}, we find that the
amplification factor is $A \sim \epsilon^2 \nu_\ast^{-2}$ for
$\epsilon^{-1} \nu_\ast \ll 1$, and $A = 1$ for $\epsilon^{-1}
\nu_\ast \gg 1$.

There are no ripple wells formed by a perturbation with $L \nabla \ln
B_1 \sim 1$, but the addition of the perturbation $\epsilon B_1$
changes the height and position of the minima and maxima. This effect
is studied in \cite{calvo13} for stellarators close to quasisymmetry,
where it is shown to be a higher order effect. The estimation for
stellarators close to omnigeneity is very similar, and gives the same
result.

\emph{Conclusions.} We summarize our results in
Fig.~\ref{fig:amplification} where we sketch the dependence on
$\nu_\ast$ of the amplification of the energy flux $A$ due to
deviations from omnigeneity.

For deviations with large
  gradients, the amplification is considerable at small
  collisionalities $\nu_\ast \ll \epsilon^{1/2}$. In this regime the
transport is dominated by particles trapped in the wells of the
omnigeneous piece of the magnetic field. Importantly, ripple wells are
not crucial for this type of transport, as has sometimes been
assumed. This assumption is usually based on the incorrect impression
that seminal work like \cite{ho87} applies to stellarators close to
omnigeneity. Unlike in \cite{ho87}, the number of particles in ripple
wells is not small because these particles do not have in general
small $v_{||}$. The particles that get into ripple wells by collisions
come from a population that has $v_{||} \sim v_{ti}$. This is the
reason why, when we assume an $O(\epsilon^{-2})$ number of wells, we
obtain the flux $\epsilon \nu_\ast^{-1} \rho_\ast^2 p_i v_{ti} S_r$
instead of $\epsilon^{3/2} \nu_\ast^{-1} \rho_\ast^2 p_i v_{ti} S_r$.

Surprisingly, due to the large gradients associated with the deviations from omnigeneity, the energy flux is unlikely to depend quadratically on the deviations from omnigeneity even for relatively small deviations. The dependence will be between linear and quadratic, and this fact will necessarily affect the competition between different optimization criteria.

\begin{acknowledgments}
This work was supported by
  EURATOM and carried out within the framework of the EUROfusion
  Consortium. This project has received funding from the EU Horizon
  2020 research and innovation programme. The views and opinions
  expressed herein do not necessarily reflect those of the European
  Commission. This research was supported in part by grant
  ENE2012-30832, Ministerio de Econom\'{\i}a y Competitividad, Spain.
\end{acknowledgments}


\begin{thebibliography}{10}

\bibitem{helander12}
P. Helander et al., Plasma Phys. Control. Fusion {\bf 54}, 124009 (2012).

\bibitem{wobig93}
H. Wobig, Plasma Phys. Control. Fusion {\bf 35}, 903 (1993).

\bibitem{nuhrenberg95}
J. N\"uhrenberg et al, Trans. Fusion Technology {\bf 27}, 71 (1995).

\bibitem{anderson95}
F.S.B. Anderson et al., Fusion Technol. {\bf 27}, 273 (1995).

\bibitem{neilson02}
G.H. Neilson et al, J. Plasma Fusion Res. {\bf 78}, 214 (2002).

\bibitem{cary97a}
J.R. Cary and S.G. Shasharina, Phys. Rev. Lett. {\bf 78}, 674 (1997).

\bibitem{cary97b}
J.R. Cary and S.G. Shasharina, Phys. Plasmas {\bf 4}, 3323 (1997).

\bibitem{garren91b}
D.A. Garren and A.H. Boozer, Phys. Fluids B {\bf 3}, 2822 (1991).

\bibitem{hazeltine73}
R.D. Hazeltine, Plasma Phys. {\bf 15}, 77 (1973).

\bibitem{northrop63}
T.G. Northrop, \emph{The Adiabatic Motion of Charged Particles}, John Wiley, New York (1963).

\bibitem{boozer04}
A.H. Boozer, Rev. Mod. Phys. {\bf 76}, 1071 (2004).

\bibitem{landreman12}
M. Landreman and P.J. Catto, Phys. Plasmas {\bf 19}, 056103 (2012).

\bibitem{parra14}
F.I. Parra, I. Calvo, J.L. Velasco and J.A. Alonso, "Stellarators close to omnigeneity", in preparation.

\bibitem{calvo13}
I. Calvo, F.I. Parra, J.L. Velasco and J.A. Alonso, Plasma Phys. Control. Fusion {\bf 55}, 125014 (2013).

\bibitem{ho87}
D.D.-M. Ho and R.M. Kulsrud, Phys. Plasmas {\bf 30}, 442 (1987).

\end{thebibliography}
\end{document}